\documentstyle[12pt, aaspp4]{article}

\begin{document}

\title{Stellar populations in the dwarf spheroidal galaxy Leo I}

\author{Filippina Caputo\altaffilmark{1}, Santi Cassisi\altaffilmark{2}, 
Marco Castellani\altaffilmark{3}, Gianni Marconi\altaffilmark{3} \and 
Patrizia Santolamazza\altaffilmark{1}}
\lefthead{Caputo et al.}
\righthead{Stellar populations in Leo I}

\altaffiltext{1}{Osservatorio Astronomico di Capodimonte, Via Moiariello 16,
80131 Napoli, Italy; caputo@astrna.na.astro.it, patrizia@cerere.na.astro.it}
\altaffiltext{2}{Osservatorio Astronomico di Collurania, Via Maggini, 64100 
Teramo, Italy; cassisi@astrte.te.astro.it} 
\altaffiltext{3}{Osservatorio Astronomico di Roma, Via Frascati 33, 00040 
Monteporzio; mkast@coma.mporzio.astro.it, marconi@coma.mporzio.astro.it}

\begin{abstract}

We present a detailed study of the color magnitude diagram (CMD) 
of the dwarf 
spheroidal galaxy Leo I, based on archival {\it Hubble Space Telescope} data. 
Our photometric analysis, confirming previous results on the brighter portion 
of the CMD, allow us to obtain an accurate sampling of the stellar 
populations also at the faint magnitudes corresponding to the Main Sequence. 
By adopting a homogeneous and consistent theoretical scenario for both 
hydrogen and central helium-burning evolutionary phases, 
the various features observed in 
the CMD are interpreted and 
reliable estimations for both the distance modulus 
and the age(s) for the main stellar components of Leo I are derived. 
More in details, from the upper luminosity of the Red Giant Branch and the 
lower luminosity of the 
Subgiant Branch we simultaneously constrain 
the galaxy distance and the age of the oldest stellar population in Leo I.  
In this way we obtain a distance modulus $(m-M)_V$=22.00$\pm$0.15 mag and an 
age of 10--15 Gyr or 9--13 Gyr, adopting a metallicity $Z$=0.0001 and 0.0004, 
respectively. The reliability of 
this distance modulus has been tested by comparing 
the observed distribution of the Leo I anomalous Cepheids in the 
period-magnitude diagram with the predicted boundaries of the instability 
strip, as given by convective pulsating models. 
The detailed investigation of the age (s) of the Leo I stellar 
populations is then performed by comparing the CMD 
with a suitable set of theoretical isochrones and central helium-burning 
models. By taking into account all the various features, including the 
lack of RR Lyrae variables, we conclude 
that the star formation process in Leo I has started at $\sim$  
10 Gyr (with $Z$=0.0001) or $\sim$ 13 Gyr (with $Z$=0.0004) 
ago and it stopped about 1 Gyr ago. Some evidence is reported supporting 
the mild metal deficiency ($Z$=0.0004), whereas no clear 
indication has been found supporting
a star formation history characterized by episodic bursts. 
The adoption of updated physics which 
includes the inward diffusion of elements, as 
recently presented for globular cluster stars, 
would yield a slightly larger distance modulus ($\sim$ 0.10 mag)  
and a slightly lower age for the most ancient stellar component 
($\sim$ 1 Gyr). 

\end{abstract}


\pagebreak
\section{Introduction}

Many of the nine dwarf spheroidal galaxies (dSphs) 
clustering around the Milky Way are 
known to have had a complicated evolutionary 
history as suggested from 
clear evidences of star formation, continuously or in bursts, 
over a wide period of time (see, e.g., Mighell 1997 [Carina],  
Beauchamp et al. 1995 [Fornax],  Lee et al. 1993 [Leo I], 
Mighell \& Rich 1996 [Leo II], Da Costa 1984 [Sculptor], and the 
comprehensive review by Mateo 1998). 

For the not--too-far dSphs, several CMDs 
that reached the main--sequence turnoff (MSTO) have been published, 
allowing the analysis of their stellar content. 
For the more distant galaxies, only in the very recent 
time the 
{\it Hubble Space Telescope} 
is providing the deep CMDs (see, e. g., Mighell \& Rich 1996)
that are necessary to fully understand the
evolutionary history of these faint members of the Local Group (van den Bergh
1994). 

In this work we present a study of the stellar populations in the 
dSph Leo I,  
based on archival 
Wide Field Planetary Camera 2
(WFPC2) data. This galaxy, discovered by Harrington \& 
Wilson (1950) during the first Palomar sky survey, is thought to be among 
the most distant satellites of the Milky Way and therefore it plays an 
important role in determining the mass of our galaxy (Zaritsky et al. 1989; 
Zaritsky 1991). On the other hand, the variable star survey carried out 
by Hodge \& Wright (1978) showed an unusual large number 
of anomalous Cepheids, and the CMD published by Lee et al. (1993 [L93]) 
"...shows no suggestion for any Horizontal Branch typical of other 
dSphs". Furthermore, it should be added that all the published 
CMDs (see also Fox \& Pritchet 1987; Reid \& Mould 1991; Demers, Irwin \& 
Gambu 1994 [DIG]) {\it suggest} that the stars of Leo I 
have a younger mean age than that of the other dSphs, but the so far  
published data - even those 
from the deepest CCD photometry - do not reach the faint magnitudes   
we need to estimate definitively the stellar age(s). 

Section 2 deals with the observations and data reduction. The CMD for Leo I 
is discussed in Section 3, together with a review of former works on this galaxy. 
The theoretical scenario used for interpreting the stellar content of Leo I 
is presented in Section 4, while our original analysis, with a discussion of the 
resultant distance and age, is given in Section 5. The summary of the main results  
follows in Section 6. 

\section{Observations and data reduction}

The data for Leo I have been requested and retrieved 
electronically  from the ESO/ST-ECF archive in Garching (Munchen). 
The galaxy was observed with the {\it Hubble Space Telescope} WFPC2 on 1994 
March 5 through the F555W ($\sim V$) and F814W ($\sim I$) filters. 
The WFPC2 aperture was 
centered on the target position $\alpha_{2000}$ = $10^h$ $08^m$ $26.58^s$, 
$\delta_{2000}$ = $12^{o}$ $18^{'}$ $33.4^{''}$, 
and eight observations were 
obtained: three 1900 s plus one 350 s exposures in F555W, and three 
1600 s plus one 300 s exposures in F814W. 
These observations (part of the HST Cycle 4 program GTO/WFC 
5350) were placed in the public 
data archive on 1995 March 5. 

Correction to the raw data for bias, dark and flat-field 
were performed using the standard HST pipeline.
Subsequent data reduction, for the WF cameras, was made using MIDAS routines
ROMAFOT and DAOPHOT II packages. 
The next steps were schematically as follows.
First, a filter median was applied upon each frame, using a software in ROMAFOT
in order to remove cosmic rays from single frames.

To push the star detection limit to as faint a level as possible, 
we coadded all images taken through the same filter. 
Then, we used the deepest I coadded frame to search for stellar objects in each
chip. All the objects identified in this search were fitted in all frames I and
V using DAOPHOT II and
the hybrid weighted technique described by Cool and King (1995). 
Two DAOPHOT detection 
passes were carried out, separated by photometry and subtraction of all stars 
found in the first pass. Faint stars hidden within the PSF skirts of 
brighter companions were thereby revealed and added to the list of detected 
stars to be measured. For each chip and each filter the PSF was build by 
using not less than 15 bright and isolated stars in each frame. 
The single measures were averaged, and an average instrumental magnitude
was derived for each object in each colour.

A total of 36.634 stars were ultimately detected and measured in the 
three WFs frames. 
Corrections to $0.5^{''}$ aperture were made in each case; we transformed the 
F814W and F555W instrumental magnitudes into the  WFPC2 
``ground system'' using Eq. 6 of Holtzman et al. (1995).

Completeness tests were carried out by adding 5$\%$ of the original number 
of stars for each selected bin (0.4 mag) of magnitudes to 
the original coadded F555W and F814W frames. 
\lq{Artificial}\rq\ stars were added randomly with 
the same instrumental color distribution of
the real stars detected in the frames.
The \lq{artificial}\rq\ frames were then 
processed using DAOPHOT II in a manner identical to that applied  to the 
original data. The completeness was finally derived as the ratio 
$N_{rec}/N_{add}$ of the artificial stars generated. We 
have considered as recovered only those stars which have been 
found in the same spatial position and inside the same magnitude bin with
respect 
to the added stars. 
The results of these tests are shown in Table 1, which shows that the 
50$\%$ completeness level occurs at F555W $\sim$ 26.0.
Since we only perform qualitative analysis, i.e., we do not make comparisons
between observative/theoretical luminosity functions, 
such a procedure is fully satisfactory for the present investigation.

The internal errors were also estimated computing the rms frame to frame 
scatter of the instrumental magnitudes obtained for each stars 
(see Table 2). 

Recently, the Leo I galaxy has been also studied by Gallart et al. (1999). They 
choose to normalize their HST photometry to L93 calibration. According to the 
quoted authors, the L93 photometry is regarded as reliable,
because the large number of calibration stars in their field.
With the aim to check qualitatively
our photometry, we made the colours distribution histogram
of the objects belonging to the clump of stars near the Red Giant Branch with 
magnitude in the range 21.5$\le V \le$22.6 mag (see Fig. 1 and the following 
discussion); a comparison with the
same histogram realized by L93 (see their Figure 8) shows 
that the peaks of the two distributions fall
in the same color bin (i.e., $ 0.80 \leq (V-I) \leq 0.85 $). 
Such result make us quite confident on the reliability of our photometry: 
differences between our calibration and that of L93 are inside the adopted
bin width (i.e., $\Delta{(V-I)}=0.05$). For a detailed star-by-star comparison, 
we address the interested reader to Gallart et al. (1999).

% ----------------------------------------------------------------

\section{CMD of Leo I}

As already stated, this section deals with the main features of the CMD of
Leo I stars as well as with a review of previous works. In our belief, this would
provide the best framework for our results, presented in the following sections.

\subsection{General morphology}

The $V$-$(V-I)$ 
color-magnitude diagram of Leo I,  as based on 
36.634 stars down to a limiting magnitude of 
$V\sim$27.5 mag ($I\sim$26.5 mag) is displayed in Fig. 1. 
The principal features,  
which are discussed 
in detail in the following section, are here summarized. 

{\it Red Giant Branch}. There is a well-defined red giant branch (RGB) 
with the tip (TRGB) seen at 
$V_{TRGB} \sim$19.4$\pm 0.10$ mag ($I_{TRGB} \sim$18.0$\pm 0.10$ mag), 
in agreement with 
the L93 study. The few stars located above 
the TRGB  
are likely to belong to the 
asymptotic giant branch (see L93 and DIG).
The observed color dispersion read 
at $V$=20.0 mag ($\simeq$ 0.5 mag below TRGB) 
is $\Delta (V-I)\sim$0.10 mag, which is slightly larger than 
the L93 value (0.08 mag). It is known that the color dispersion along RGB 
can derive from photometric 
errors as well as from metallicity and age spread. However, 
at $V$=20.0 mag the mean photometric error (see Table 2) is   
$\sim$ 0.015 mag, leaving an intrinsic dispersion of 
(0.10$^2$-0.015$^2$)$^{0.5}$ = 0.098 mag which will be discussed 
in the following. 

{\it Horizontal Branch}. As already shown from previous studies, 
there is no evidence of the "flat" portion of the horizontal  
branch (HB) which is typical of Galactic globular clusters and 
other dSphs. However, the clump of red giant stars 
seen at $(V-I)\sim$ 0.7-0.9 mag and  
21.5$\le V \le$22.6 mag are {\it bona fide} 
central helium--burning stars, even though 
more massive  
than those observed in old stellar systems. 
The sequence of stars with 20.0$\le V \le$21.5 and 
0$\le (V-I) \le$0.7 could represent the 
more massive tail of these helium-burning stars (see Section 4). 

{\it Main Sequence stars}. The most impressive 
feature in the color-magnitude 
diagram of Leo I  is the MSTO region seen 
at {\it V}$\sim$
22.60$\pm$0.20 mag, which is also the luminosity of the faintest 
helium-burning clumping stars (the lower envelope 
of HB stars is taken at $V_{HBLE}=22.60\pm$0.05). Such a feature 
would suggest by itself the presence of
stars with age near 1--2 Gyr 
(see Caputo \& Degl'Innocenti 1995). Moreover,  
there is also a small group of brighter main--sequence stars with 
21.0$\le V \le$22.6 mag. These "blue stragglers" could be mass transfer 
binaries or, if normal main sequence stars, they might be 
witnesses of a small population of very young stars.  
Beside the above clear evidence of a young 
stellar component, one notices the well developed   
subgiant branch extending below the clumping red giant stars. The faint stars 
at the base of the subgiant branch (BSGB) are seen at 
$(V-I)\sim$0.8 mag and 
$V_{BSGB}=25.00\pm$ 0.20 mag. The observed difference in 
magnitude between the TRGB and BSGB stars 
is $\Delta V(BSGB-TRGB)\sim$ 5.6 mag, which is similar to the 
values observed in the CMDs of Galactic globular clusters, thus suggesting 
an old stellar 
population of $\sim$ 10--15 Gyr.

\subsection{Variable stars}

As stated at the very beginning,  
the most striking difference between Leo I and the other dwarf spheroidals 
is the lack of RR Lyrae stars and, conversely, the large number of anomalous 
Cepheids. Hodge \& Wright (1978) measured 
blue magnitude, period and amplitude for 
12 variables (with an estimate of 75\% completeness), 
while L93 suggested that the 45 stars observed with 
21.2 $\le V \le$ 19 
mag and 0$\le (V-I) \le$ 0.6 mag are anomalous 
Cepheid candidates. None of the 
Hodge \& Wright (1978) variables are located in our color-magnitude diagram 
and for the stars in our CMD seen with 
20.0$\le V \le$ 21.5 and
0 $\le (V-I)\le$0.6  we have no way to 
confirm the variability.

\subsection{Metallicity and reddening}

There are somewhat conflicting results concerning the mean metallicity 
of Leo I. Previous estimates based on CMD features 
vary from [Fe/H]=-1.0$\pm$0.3 
(Reid \& Mould 1991), to [Fe/H]=-1.6$\pm$0.4 [DIG] and   
[Fe/H]=-2.1$\pm$0.1 [L93], depending on the assumed distance 
modulus, while moderate resolution 
spectra of two red giants (Suntzeff 1992, see L93) 
suggest [Fe/H]$\sim$-1.8. 
On the other hand, we show in Section 4 that 
the occurrence of a significant number of 
anomalous Cepheids is by itself a clear indication that Leo I is 
a metal--poor stellar system  
with an overall metallicity between $Z$= 0.0001 and 0.0004 (see 
also Castellani \& 
Degl'Innocenti 1995; Caputo \& Degl'Innocenti 1995; Bono et al. 1997 
[BCSCP]).

As for the reddening, the relatively high galactic latitude of Leo I 
suggest a low foreground reddening. The blue extinction reported by 
Burstein \& Heiles (1984) is $A_B$=0.09 mag. On this basis, following 
the Cardelli, Clayton \& Mathis (1989)
relations, we will adopt $E(B-V)$=0.02 mag and $E(V-I)$=0.04 mag. 

\subsection{Distance and Age}

Given the lack of HB stars at the RR Lyrae gap, 
previous estimates of the Leo I distance 
have mostly used the TRGB [L93: $(m-M)_0=22.18\pm$0.11 mag], the median 
magnitude of the red giant clump 
[DIG: $(m-M)_0=21.7\pm$0.12 mag] and the carbon stars 
[DIG: $(m-M)_0=21.5\pm$0.3 mag]. As early 
discussed by DIG, the problem with these 
distance indicators is that the results depend on the adopted 
metallicity (see also Cassisi, Castellani \& Straniero 1994) 
in the sense that the deduced distance modulus increases with 
decreasing the metal content. We wish to add that some of the theoretical 
constraints discussed in this paper suggest a  
dependence of the TRGB luminosity on the age which has to be 
taken into account when dealing with 
young stellar populations.  

As for the age, the absence of blue horizontal branch, the presence 
of several carbon stars, and the clumping 
red giant stars yielded DIG to suggest an upper age limit 
of $\sim$ 7 Gyr for the dominant stellar 
population, with no obvious evidence of an older stellar 
component (their CMD does not reach the main sequence turnoff). The deeper 
CCD photometry presented by L93 revealed 
an increased number of main sequence stars 
at $V\sim$ 23.5 mag, consistent with the presence of young stars of 
$\sim$ 3 Gyr. More recently, the L93 measurements have been interpreted 
by Caputo, Castellani \& Degl'Innocenti (1995, 1996) as 
evidence of even younger stellar populations ($\sim$ 2.0--1.5 Gyr). 

\section{Theoretical background}

In order to provide a clear and complete 
reference framework, let us summarize the 
primary theoretical constraints which are relevant for the 
present investigation. They are derived from the evolutionary models 
(both hydrogen and central helium-burning phases) 
with masses from 0.6$M_{\odot}$ 
to 2.2$M_{\odot}$, original helium $Y_0$=0.23 and metallicity $Z$=0.0001, 
0.0004 already presented by Castellani 
\& Degl'Innocenti (1995) and BCSCP. 
\par
As a first point, we show in Fig. 2 
theoretical isochrones with $Z$=0.0001 and selected 
ages, transformed 
into the observational plane $M_V$-$(V-I)$ by means of the stellar 
atmosphere models provided by Castelli, Gratton \& Kurucz (1997a,b). 
Besides the well known evidence that with increasing age  
the RGB color becomes redder and the MSTO point 
becomes fainter, one notices the clear variation of 
the TRGB luminosity, which fades at the 
lower ages. As shown in the lower panel of Fig. 3, where
the predicted absolute magnitude of TRGB is plotted versus the age, 
such a behaviour is less pronounced with $Z$=0.0004. 
For the purpose of present paper, we present in  
the upper panel of Fig. 3 the corresponding variation 
of the luminosity at the base of the subgiant branch. 

From the theoretical isochrones with $Z$=0.0001 
we derive that at $M_V$=-2.0 mag ($\sim$ 0.5 mag below TRGB) the 
color variation with age is 
$$(V-I)_{M_V=-2}=1.13+0.11logt,\eqno(1)$$
while the $Z$=0.0004 isochrones are redder by
$\Delta (V-I)\sim$0.04 mag, at constant age. 
Such a result yields that the predicted 
contribution of the age to the observed color dispersion along RGB is 
significantly larger than previously adopted. As an example, L93 assumes a 
difference $\Delta (V-I)$=0.03 mag between the 3.5 and 15 Gyr isochrones, 
whereas the present results would give $\sim$0.07 mag. 
On these grounds, one understands that it is necessary  to estimate the actual 
age spread before deriving the metallicity dispersion from the observed RGB 
width. 

Passing to the central helium-burning phase, let 
us remind that for stellar structures experiencing
a strong He-flash, i.e. for evolving masses smaller than $\approx1.0M_\odot$, 
the age $t_{fl}$ and the mass $M_{c,fl}$ of the He-core at the RGB tip depend 
slightly on the evolutionary mass, but significantly on the chemical composition. 
On the contrary, when increasing the stellar mass, these evolutionary parameters 
strongly depend on both the mass (as a consequence of the changes in the 
electron degeneracy level inside the core during the RGB evolution) and chemical 
composition. The data listed in Table 3 and plotted in Fig. 4 show that, if the 
mass $M_{pr}$ of the star is lower than $\approx2.2M_{\odot}$, then both 
$t_{fl}$ and $M_{c,fl}$ are decreasing functions of $M_{pr}$. 
The consequences on the subsequent zero age horizontal branch (ZAHB) are easily 
understandable (see also Castellani \& Degl'Innocenti 1995; Caputo \& 
Degl'Innocenti 1995). With increasing the age, the maximum permitted mass 
$M_{HB,max}$ for central He-burning stars ($M_{HB,max}$ is equal to 
the mass of the RGB progenitor $M_{pr}$ in the hypothesis of no mass--loss 
during the RGB phase) decreases, whereas, following the corresponding variation 
of $M_{c,fl}$, the luminosity of the ZAHB model at $(V-I)_0\sim$ 0.70 mag tends 
to increase (see last column in Table 3).

On the other hand, the evolutionary calculations show that 
the effective temperature of a ZAHB model decreases
with increasing the mass, reaching the minimum value of $\log{T_e}\sim$3.74
($Z$=0.0001) or $\sim$ 3.72 ($Z$=0.0004) around
1.0-1.2 $M_{\odot}$. After that, the more massive 
models with $Z$=0.0001 present higher luminosity and
larger effective temperature, causing a ZAHB "turn-over" and the development of 
a "upper horizontal branch" (UHB). On the contrary, the models with $Z$=0.0004 
and mass in the range of 1.3$M_{\odot}$ to 1.5$M_{\odot}$ are    
characterized by higher luminosity and roughly constant effective temperature. 
Consequently, with $Z$=0.0004 the ZAHB "turnover" is occurring after
1.5$M_{\odot}$.

All these features are presented in Figures 5 and 6,
where ZAHB sequences (solid line) of stars with the same RGB progenitor (see the labelled
$M_{pr}$) but having experienced different degrees of mass-loss, are displayed.
The same figures show the post--ZAHB evolution of the most massive 
($M_{HB,max}$=$M_{pr}$) central He--burning model (dashed line).
Note that a further increase of the metallicity up to $Z$=0.001 
would shift the ZAHB "turn-over" to masses significantly 
larger than 2.0$M_{\odot}$ (Demarque \& Hirshfeld 1975; Hirshfeld 1980). 

In order to have an immediate insight into the connection
between the central--helium burning evolution and radial pulsation, 
we show in Figures 7a and 7b the evolution of HB models with mass 
$M_{HB}$=$M_{pr}$, but with  the effective temperature of the model
scaled to the red edge of the instability strip (FRE). 
The location of FRE, as well as the
adopted width of the instability strip, is provided from
the pulsational models discussed by BCSCP.

The first straightforward result is that He-burning stars with evolutionary 
mass in the range of $\sim 1.0M_{\odot}$ to $\sim 1.2M_{\odot}$ (for $Z$=0.0001)
or in the range of $\sim 0.8M_{\odot}$ to $\sim 1.7M_{\odot}$ (for $Z$=0.0004)
are confined near the red giant branch, out of the
instability strip. Thus, no variable stars are expected 
within these mass ranges. Moreover, one derives that the lowest 
mass for the occurrence of massive pulsators brighter that RR Lyrae 
stars, i.e. anomalous Cepheids, is 1.3$M_{\odot}$ with $Z=$0.0001 and
1.8$M_{\odot}$ with $Z$=0.0004. In terms of age, this could mean that 
the anomalous Cepheids should have ages younger than $\sim$ 3 and 1 Gyr, 
with $Z$=0.0001 and $Z$=0.0004, respectively (see Table 3). Finally, 
one may notice that with $t_{fl}\sim$ 15 Gyr the evolution of 
the most massive HB model 
at $Z$=0.0004 ($M_{HB}=M_{pr}=0.75M_{\odot}$) is confined at the 
red side of the RR Lyrae instability strip, whereas  with $Z$=0.0001 
the most massive HB model (0.80$M_{\odot}$) 
evolve within the instability strip. Thus, 
{\it even with a null mass-loss}, central He-burning         
stars with $Z$=0.0001 and age $\sim$ 
15 Gyr are expected to populate the RR Lyrae gap. 

\section{Revising the distance modulus and the age 
of Leo I}

The theoretical isochrones presented in Fig. 2 show that with increasing 
age the maximum luminosity of RGB stars increases whereas the minimum 
luminosity of SGB stars decreases. Thus, if composite stellar populations 
are present in Leo I, then the oldest stars 
have to be seen at BSGB and TRGB. By starting from these simple 
considerations, we combine in Fig. 8 the theoretical data already shown in 
Fig. 3 with the observed values $V_{TRGB}=19.40\pm$ 0.10 mag
and  $V_{BSGB}=25.00\pm$ 0.20 mag, aiming at checking the possibility of a 
unique solution for the Leo I distance modulus by using these two 
observables. As a result, we obtain that the apparent distance modulus 
of Leo I, as given from its oldest stellar component, is $(m-M)_V$=22.00$\pm$ 
0.15 mag. 
Moreover, the data in Fig. 8 suggest that 
the age of these stars is in the range of 10.0--15.0 Gyr and 9.0--13.0 Gyr, 
with $Z$=0.0001 and $Z$=0.0004, respectively. However, if further observations will 
confirm the lack of RR Lyrae stars, then from the data plotted in Fig. 7a 
we could add that the Leo I oldest stellar component cannot be older 
than $\sim$ 10 Gyr, with $Z$=0.0001. 

We can straightway check the derived 
distance modulus $(m-M)_V$=22.00$\pm$ 0.15 mag 
by comparing observed data of Leo I 
anomalous Cepheids with the theoretical predictions given by  
the BCSCP pulsating convective models. Figure 9 shows the 
period--luminosity diagram for Leo I variables\footnote{The $B$-magnitudes 
from Hodge \& Wright (1978) are corrected with 
$(m-M)_B$=22.02, according to the adopted reddening 
$E(B-V)$=0.02 mag.}. It is quite evident that 
the predictions conform very well to the observed data, supporting the 
above distance modulus.   

The comparison between theoretical isochrones and the CMD of the stars in 
Leo I, corrected with $(m-M)_V$=22.00 mag and $E(V-I)$=0.04 mag, 
is displayed in Fig. 10a ($Z$=0.0001) and Fig. 10b ($Z$=0.0004). 
As a whole, these figures provide further support to the result that the 
oldest stars in Leo I were formed $\sim$ 10 Gyr or $\sim$ 13 Gyr  
(with $Z$=0.0001 and 0.0004, respectively) ago. Moreover, 
the brightest MSTO stars seen at $V\sim 22.60\pm 0.20$ mag conform quite 
well the 1 Gyr isochrones, while the bright blue stragglers should
have even younger ages ($\sim$ 700 Myr). The absence of distinct MSTOs,
as those seen in Carina (see Smecker--Hane et al. 1996),
gives evidence against episodic bursts and, as a whole,
we conclude that Leo I has forming stars rather continuously, even though
at lower level during the last billion years, from about 
10 Gyr or 13 Gyr ago (depending on the 
adopted metallicity) to at least $\sim$ 1 Gyr ago. Our conclusions are 
not discordant with the results of Gallart et al. (1998) which 
suggest that Leo I experienced a major increase of star formation from 
$\sim$ 6 to 2 Gyr ago, with some prior episodes lasting 2-3 Gyr and a 
decreasing activity until 500-200 Myr ago. 

The derived spread of ages leads [see Eq. (1)] to a predicted 
color dispersion 
along RGB of $\sim$ 0.11 mag, which is consistent with 
the intrinsic RGB width ($\sim$0.10 mag). This result seems to 
exclude a substantial metallicity dispersion of the Leo I stars. 
On the other hand, it has been shown that the presence of anomalous 
Cepheids is a clear indication of young ($\le$ 3 Gyr) {\it and} metal--poor 
($Z\le$ 0.0004) stellar population. Thus, 
we are able to conclude that the actual metal dispersion  
content of Leo I is at the most in the range of $Z$=0.0001 to $Z$=0.0004.

It has been shown (see, e.g., Caputo \& Degl'Innocenti 1995) that 
in stellar systems in which is present a {\sl not-too-old} stellar population, 
the observed star distribution along the RG clump can provide safe 
constraints on the allowed range of stellar ages. Now we wish 
to adopt a similar approach to investigate how the above age dispersion
agrees with the observed clump of central helium-burning stars. For this aim, 
the CMD of Leo I stars with $V\le$ 23.00 mag 
is displayed in Fig. 11 together with HB 
evolutionary tracks with $Z$=0.0001 and $M_{HB}=M_{pr}$ 
(the dashed line refers to the model with 
2.0$M_{\odot}$), and by adopting the two extreme values of the distance 
modulus derived from the previous analysis. 
With $(m-M)_V$=21.85 mag (lower panel), the absolute magnitude of the
lower envelope is equal to $M_V^{HBLE}$=0.75$\pm$0.05 mag, suggesting 
(see Table 3) that the stars at the HBLE are $\sim$ 2 Gyr old and have a RGB 
progenitor with mass $M_{pr}\sim 1.4M_\odot$. However, when considering the 
mass distribution along the corresponding ZAHB locus, 
the mass of the star at $(V-I)_0=0.7$ mag is of about 
0.8$M_{\odot}$, a result which would imply a substantial mass-loss during the RGB 
phase (or at the He-flash). 

As for the remaining stars forming the clump seen at $(V-I)\sim$ 0.7-0.9 mag and
21.5$\le V \le$22.6 mag, 
the comparison with our HB evolutionary models shows that they 
are matched by the HB evolution of models with mass ($M_{HB}=M_{pr}$) equal to
0.9$M_{\odot}$ ($\sim$ 10 Gyr), 1.0$M_{\odot}$ ($\sim$ 7 Gyr) and 
1.2$M_{\odot}$ ($\sim$ 4 Gyr). Similarly, 
the stars with 20.0$\le V \le$21.5 and
0$\le (V-I) \le$0.7, which observationally define the UHB, appear 
reasonably fitted with more massive (and younger) HB models from 1.4$M_{\odot}$ 
(2.2 Gyr) up to $\sim 2.0M_{\odot}$ ($\sim$ 0.7 Gyr), assuming no mass-loss.
Let us notice, that we are not 
neglecting the possibility that a mass-loss phenomenon could affect the 
progenitors of such HB structures during the RGB evolution. Here we are 
interested only to the location in the CMD of the more massive (and brightest) HB star 
for each fixed assumption on the RGB progenitor. All the other less massive ZAHB
structures are located at lower luminosity (see previous 
discussions and Caputo \& Degl'Innocenti 1995). For the same reason, in principle it 
could be possible that such CMD region is populated by HB structures with still 
more massive - and then younger - RGB progenitor, which have suffered an 
efficient mass-loss phenomenon during the previous evolutionary phase. 
Nevertheless, this occurrence does not seem supported at all by the comparison 
between the full CMD diagram and theoretical isochrones, performed in Fig. 10.

By adopting $(m-M)_V$=22.15 mag (upper panel),
the absolute magnitude of the HBLE stars 
$M_V^{HBLE}$=0.45$\pm$0.05 mag, which is consistent with 
the location of the 10 Gyr old ZAHB (corresponding to the 0.9$M_\odot$ 
progenitor) and masses (on the ZAHB) near 0.9$M_{\odot}$, thus implying a 
negligible mass-loss. 
As a whole, for the clumping red giant stars with 
$(V-I)\sim$ 0.7-0.9 mag and 21.5$\le V \le$22.6 mag we derive 
masses (during the central He-burning
phase) from 0.9 to 1.3$M_{\odot}$ and ages   
in the range of 10 to 3 Gyr, respectively, assuming no mass-loss for the RGB 
progenitor. 
As for the UHB stars with 
20.0$\le V \le$21.5 and 0$\le (V-I) \le$0.7, they appear somehow 
brighter than the 
evolutionary tracks of the most massive (and younger) models, 
rather supporting the smaller distance modulus. 
In passing, we wish to note the fine agreement between the shape of the observed 
clump of stars and the location of our HB tracks.
However, we notice also some discrepancy between observed
data and theoretical models with $Z$=0.0001 as due to the blue loop
of the evolutionary sequences which extends
hotter than the observed color of the clumping stars.

As shown in Fig. 12, such a discrepancy is removed if the $Z$=0.0004 models 
are taken into consideration. By adopting $(m-M)_V$=21.85 mag (lower panel), 
the HBLE stars turn out to be $\sim$ 2.3 Gyr old, with a 
mass near 0.75$M_{\odot}$ to be compared with $M_{pr}$=1.4$M_{\odot}$. 
On the other hand, the remaining clumping red giants 
with $(V-I)\sim$ 0.7-0.9 mag and 21.5$\le V \le$22.6 mag agree with the 
HB evolution of models from 0.80 to 1.4$M_{\odot}$ and 
ages from 13 to 2.3 Gyr, respectively, assuming no mass-loss. Similarly, 
the brightest stars with 20.0$\le V \le$21.5 and 0$\le (V-I) \le$0.7
seem to require masses up to 2.2$M_{\odot}$ (dotted line), assuming no 
mass-loss for the progenitor. Adopting 
$(m-M)_V$=22.15 mag (upper panel) yields that the absolute magnitude of the
HBLE 
stars is somehow brighter than the 0.80$M_{\odot}$ model with age 
of $\sim$ 13 Gyr (i.e. the maximum age 
derived from isochrone fitting), 
suggesting that the distance modulus of Leo I is not larger 
than $(m-M)_V$=22.00 mag. With such a value, we derive that the HBLE 
stars have mass 0.80$M_{\odot}$ and age $\sim$ 13 Gyr, while for the remaining 
clumping red giant stars we obtain 1.0--1.6$M_{\odot}$ and 1--7 Gyr, 
respectively, assuming no mass-loss. However, 
also for this metallicity the fit 
of the UHB stars seems to support the smaller distance modulus. 

In conclusion, by taking into account all the various features of the CMD, 
our best estimates for the metallicity and the distance modulus of Leo I 
are $Z$=0.0004 and $(m-M)$=21.90$\pm$0.05 mag. The resulting ages of the
stellar 
components are from 1 to 13 Gyr, with few stars as young as $\sim$ 700 Myr, 
as derived from theoretical isochrones and HB evolutionary models. 
Finally the analysis of the clumping red giants seem to suggest that 
the younger stellar populations (i.e. those with massive RGB progenitors) 
suffered a substantial mass-loss during the RGB phase or at the He-flash. 

Before closing this analysis, it seems worth mentioning that the recent 
improvements of the stellar evolutionary models have produced younger ages 
for the Galactic globular clusters (see, e.g., Cassisi et al. 1998 and 
references therein). Even though the whole "improved physics" is subject 
of deep investigation (Castellani \& Degl'Innocenti 1998), we decided 
to compute a set of "new" evolutionary models with $Z$=0.0002, 
taking into account also the inward diffusion of helium and heavy elements.  
>From the new isochrones shown in figure 13, we obtain slightly larger  
distance modulus ($(m-M)_V$=22.10$\pm$ 0.15 mag) and ages 
in the range of $\sim$ 0.7 to 10 Gyr. As for the He-burning 
stars, the new models 
yield that the red giant clumping stars have ages and masses in the range 
of $\sim$ 1 to 10 Gyr and 0.85 to 1.3$M_{\odot}$ respectively, 
while the UHB stars require masses up to 2.2 $M_{\odot}$ (see figure 14). 

\section{Summary}

As clearly stated in several works (see the comprehensive review 
by Mateo (1998) and reference 
therein) the possibility to obtain a deep insight into the intrinsic 
evolutionary properties of 
the main stellar population(s) in dSphs represents a pivotal tool in order 
to understand not only the 
properties of our nearest extragalactic neighbours but also to 
improve our knowledge of the 
Galaxy. Indeed, the decoding of the evolution history of dSphs in the 
Local Group sheds light on the formation and evolution of our own Milky Way. 
Moreover, a thorough understanding of the 
evolution history of dSphs and more in general of the Local 
Group is a fundamental precondition in order to understand the 
observational features of 
high-redshift, unresolvable galaxies. 
However, it is clear that accurate analysis of the star 
formation history and reliable knowledge of age(s) and 
metallicity distribution rely on several observational features, as 
derived from as much accurate 
as possible CMD  
diagrams reaching the faintest Main Sequence magnitudes. 
In the present analysis we have adopted this approach in order 
to investigate into the stellar populations 
of the dSph galaxy Leo I. The more relevant points 
can be summarized as it follows.

\begin{itemize}                                                           
\item	By using HST archival data, an accurate photometric investigation  
has been carried out. 
This occurrence has allowed us to obtain a CMD with more than 
36.600 stars which reaches very faint magnitudes 
($V\sim$ 27.5 mag). The tests  
performed during the photometric analysis show that our photometry 
reaches the 50\% 
completeness level at the magnitude (F555W) $\approx26.0$.
\item	The CMD of Leo I 
is characterized by 
a well-defined RGB and by a HB clumped near 
the RGB. The observed TRGB is seen at $V$=19.40$\pm$0.10 mag and 
the HB morphology does not show any evidence for a 
{\it flat} distribution near the RR Lyrae instability strip. Moreover, 
the well developed SGB extends below the red giant clump, down to 
$V$=25.00$\pm$0.20 mag and the brightest MSTO is
located at
$V\sim22.60\pm0.20$ mag, with some few stars even brighter
and bluer.
\item	By adopting a reference theoretical scenario for both hydrogen 
and central helium-burning stars with $Z$=0.0001 and $Z$=0.0004, 
the observed maximum luminosity of RGB and minimum luminosity of SGB 
are used to derive a distance modulus of $(m-M)_V$=22.00$\pm$0.15 mag. 
Furthermore, the resulting estimates for the age of the oldest 
stellar population turn out to be in the range of 10--15 Gyr 
and 9--13 Gyr with $Z$=0.0001 and 0.0004, respectively. However, 
when considering also the lack of RR Lyrae stars in Leo I, 
we conclude that the oldest stellar component in Leo I is at the most 
10 Gyr old, with $Z$=0.0001.  
\item	Such distance modulus evaluation has found further support 
by comparing the distribution of anomalous Cepheids in 
the $<M_B> - \log{P}$ plane with the location of 
the instability strip boundaries, as predicted by convective pulsating 
models.
\item	By adopting the above distance modulus, the comparison 
of the Leo I CMD with theoretical isochrones yields
that the brightest MSTO stars are consistent 
with an age of the order of 1 Gyr, for both the two adopted 
metallicities. 
\item	When all these evidences are accounted for, it is possible to reach 
the conclusion that the star formation process in Leo I has 
started at about 10 Gyr or 13 Gyr ago, depending on the adopted 
metallicity, and it stopped about 1 Gyr ago, without 
any clear evidence for the star formation occurring by single episodic 
bursts. This results appear in satisfactory agreement with the scenario
outlined by Mateo (1998, his figure 8b) and Grebel (1998); 
\item	such a dispersion of age is consistent with the mass of 
central helium--burning stars, which are derived to be in the range 
of 0.75--0.9$M_{\odot}$ to 2.0--2.2$M_{\odot}$, depending on the adopted 
metal content.
\item	The estimated age range of the main stellar components in
Leo I provides a consistent explanation for the observed color
spread along RGB, without the need
for invoking a substantial metallicity dispersion as early
suggested by L93. 
\item	Finally, the use of the theoretical evolutionary scenario 
based on a updated 
physical inputs (Cassisi et al. 1998) does not change 
in remarkable way the results achieved:  
the main effects are a slight increase of the distance 
modulus ($\sim$ 0.10 mag) and a slight  
decrease of the age for the older stellar component ($\sim$ 1 Gyr).
\end{itemize}

\acknowledgments{We are grateful to an anonymous referee for the pertinence
of her/his comments regarding the content and the style of an early draft of this
paper, which improved its readability.}

\pagebreak

\newpage

\centerline{\bf Figure captions}

\figcaption[]{The $V$ vs $(V-I)$ color magnitude diagram of 36.634 stars in Leo
I down to a limiting magnitude of $V\sim$ 27.5 mag. The photometric errors at 
different magnitudes have been also plotted (see Table 2).}

\figcaption[]{Theoretical isochrones for $Y$=0.23, $Z$=0.0001 and ages from 1 to 
10 Gyr (left to right).}

\figcaption[]{Predicted absolute magnitude at the tip of RGB (lower panel) and
at the 
base of SGB (upper panel) as a function of age and for the two selected
metallicities.}

\figcaption[]{The behavior of the age $t_{fl}$ (lower panel) and mass of the He-
core $M_{c,fl}$ (upper panel) at the He ignition as a function of the mass 
$M_{pr}$ of the red giant progenitor and for the two selected metallicities.}

\figcaption[]{ZAHB sequences (solid line) with $Z$=0.0001 of stars with the same RGB
progenitor (see the labelled masses) and different degree of mass-loss. In 
each panel, the dashed line refers to the evolution off ZAHB of the HB model with 
mass $M_{HB}=M_{pr}$.} 

\figcaption[]{As fig. 5, but for $Z$=0.0004.}

\figcaption[]{{\it Panel a)}: The H-R diagram for HB models with $Z$=0.0001 and
mass ($M_{HB}=M_{pr}$) as labelled, but with the effective temperature scaled to 
the red edge of fundamental pulsator strip. The vertical lines define the 
boundaries of the instability strip. {\it Panel b)}: as
in panel a, but for $Z$=0.0004.}

\figcaption[]{The distance modulus - age diagram for the oldest stellar component 
of Leo I, as obtained by using the observational measurements for the TRGB 
(filled circles) and BSGB (open circles) magnitudes, and the theoretical constraints
provided in Figure 3.}

\figcaption[]{The period-luminosity diagram for anomalous Cepheids in Leo I in
comparison 
with the predicted limits for pulsation. Solid and dashed lines refer to
$Z$=0.0001 and $Z$=0.0004, respectively.}

\figcaption[]{{\it Panel a)}: comparison between the CMD of Leo I and 
theoretical isochrones for $Z$=0.0001 and ages from 700 Myr to 15 Gyr. {\it 
Panel b)}: as in panel a), but with a metallicity $Z$=0.0004 and ages in the range from
1 to 15 Gyr.}

\figcaption[]{CMD of stars with $V\le$ 23.00 mag in comparison with HB models
with $Z$=0.0001 and mass $M_{HB}=M_{pr}$=0.9, 1.0, 1.2, 1.4, 1.6, 1.8$M_{\odot}$ 
(solid line) and 2.0$M_{\odot}$ (dashed line) for two different assumptions on 
the Leo I distance modulus.}

\figcaption[]{As in Fig. 11, but for HB models with $Z$=0.0004 and mass
$M_{HB}=M_{pr}$=0.75, 
1.0, 1.2, 1.4, 1.6, 1.8$M_{\odot}$ (solid line), 2.0$M_{\odot}$ (dashed line)
and 2.2$M_{\odot}$ (dotted line).}

\figcaption[]{Comparison between the CMD of Leo I with a selected set of
isochrones, 
computed by adopting updated evolutionary models (see text for more details),
for $Z$=0.0002 and for the labelled ages. The adopted values for the distance
modulus and the reddening are labelled.}

\figcaption[]{Comparison between the CMD diagram location of the HB stars in
Leo I and
updated evolutionary models (see text) for $Z$=0.0002, for two 
different assumptions on the distance modulus. The stellar masses are
$M_{HB}=M_{pr}$=0.85, 1.2, 1.5, 1.8$M_{\odot}$ (solid line) and 2.2$M_{\odot}$ 
(dotted line).}

\end{document}